\documentclass[12pt]{article}
\usepackage{graphicx}
\usepackage{amssymb,amsmath,amsfonts,amsthm}
\usepackage{epstopdf}
\usepackage[normalem]{ulem}
\usepackage{comment}
\usepackage{hyperref}

\hypersetup{
   colorlinks=true,         
   linkcolor=blue,          
   citecolor=red,        
   urlcolor=Violet            
}

\usepackage[usenames,dvipsnames]{color}

\begin{document}

\title{\Large \bf{Search for short strings in~$e^+e^-$-annihilation} }

\author{Marina~Kozhevnikova,$^1$ 
Armen~Oganesian,$^{1,2}$~~
Oleg~Teryaev,$^1$
\footnote{Electronic addresses: 
\href{mailto:armen@itep.ru,}{armen@itep.ru,} \href{mailto:teryaev@theor.jinr.ru}{teryaev@theor.jinr.ru}.}
 \vspace{12pt} \\
\it \small $^1$ Bogoliubov Laboratory of Theoretical Physics,\\
\it \small Joint Institute for Nuclear Research, 141980, Dubna, Russia,\\
\it \small  $^2$Institute of Theoretical and Experimental Physics, 117218,  Moscow, Russia.}

\date{}
\maketitle

\begin{abstract}

In this work the behavior of power corrections to Adler function in operator product expansion (OPE) is studied, in particular the possible contribution of operator of dimension 2. The OPE terms of dimension 4 and 6 are taken into account.
Various experimental data on reactions of $e^+e^-$-annihilation to pion channels ($\pi^+\pi^-$, $2\pi^+2\pi^-$, $\pi^+\pi^-2\pi^0$, $3\pi^+3\pi^-$, $2\pi^+2\pi^-2\pi^0$) are used.
The high-precision fits of the experimental data are obtained and used.
The method based on Borel transform of Adler function is applied.
It is shown that the contribution of operator of dimension 2 is negative being compatible to zero at three standard deviations level. The strong (anti)correlation between short string and local gluon condensate is found.\end{abstract}

\newpage


\section{Introduction}
\label{intro}
The problem of the existence of the operator with dimension 2, whose contribution to QCD sum rules ~\cite{SVZtf79} is proportional to~$1/Q^2$, and search for relevant OPE corrections  are performed already for quite a long time \cite{ShortString, ZakharovOther, BodDominEidel2012}. 

In the pioneering paper~\cite{ShortString} a concept of short strings leading to corrections with dimension 2 was suggested.
In Cornell potential~\cite{Eichten:1974af}
$$
V(r) \approx -\cfrac{4\alpha_s(r)}{3r}+kr
$$
the term~$kr$ describes string potential (connected to the phenomenon of confinement) at short distances, leading\footnote{Note that in OPE the first correction to $e^+e^-$-annihilation cross-section is~$\sim \left\langle G_{\mu\nu}G^{\mu\nu}\right\rangle / Q^4$.}
to the correction~$\sim k/Q^2$. In paper~\cite{ZakharovOther} the correlation between short strings and perturbation theory order was established
\footnote{It is analogous to the interplay of perturbation theory order and higher twist in deep inelastic scattering~\cite{Kataev:2000wn}}.
The contribution of the operator with dimension 2 to the $e^+e^-$ data was studied later~\cite{BodDominEidel2012, SPIESBERGER:2013jpa} and it was found to be compatible to zero with large errors.

Our analysis is based on the use of Adler function being a simple two-point
correlator which is convenient to compare to experimental data. 
We use a large number of them and perform an accurate numerical analysis. It consists of construction of the fitting model, calculating~$R$-ratio and $D$-function in dispersional form, applying Borel transform (BT), construction of sum rule and extraction of the OPE coefficients. 

The main purpose is to verify whether the operator with dimension 2 exists.





\section{Fits}
\label{sec-fits}

We limit ourselves by the data with isospin~$I=1$, thereby avoiding strange particles, which are difficult to measure. 
While the data are presented as sets of experimental points, for accurate analysis it is preferable to obtain analytic fits of experimental data which are convenient to integrate.
Whereas there are different annihilation channels data obtained from different detectors, we fit every channel separately and find the formula for full cross-section.

The analysis is performed using the data on~$e^+e^-$-annihilation to pion channels: $e^+e^-\to\pi^+ \pi^-$ (CMD and OLYA detectors)~\cite{Barkov85},
	  $e^+e^-\to 2\pi^+ 2\pi^-$ (M3N, GG2, DM1, CMD, OLYA, DM2, SND, CMD2, SND, BaBar)~\cite{M3N79,GG280,DM182, CMD88,OLYA88,DM291,SND91,CMD299,SND03,CMD204,BaBar05},
		$e^+e^-\to \pi^+ \pi^-2\pi^0$ (OLYA, CMD2, SND, DM2, Frascati-ADONE-GAMMA-GAMMA-2)~\cite{All-4pi0, OLYA4pi0, CMD24pi0,SND4pi0,ORSAY-DCI-DM2-4pi0,FRASCATI-ADONE-4pi0},
  $e^+e^-\to 3\pi^+ 3\pi^-$ (BaBar)~\cite{BaBar06},
  $e^+e^-\to 2\pi^+ 2\pi^- 2\pi^0$ (BaBar)~\cite{BaBar06}.






For description of the data on squared pion form-factor~$|F_\pi(s)|^2$ depending on energy~$\sqrt{s}$ of reaction~$e^+e^-\to \pi^+ \pi^-$ we use the three-resonance (Gounaris-Sakurai) model~\cite{Barkov85},~\cite{GouSak68}, where the form-factor of each resonance~$V$ is calculated using Breit-Wigner formula: 
$$
  F^\text{BW}(s, m_V,\Gamma_V)
   = \frac{m_V^2 (1 + d\cdot\Gamma_V/m_V)}{m_V^2-s+f(s, m_V, \Gamma_V) -i\,m_V\ \Gamma_V(s)}\,,\,\text{where} \quad \Gamma_V(s)
   = \Gamma_V\,
       \left(\frac {k(s)}{k(m_V^2)}\right)^3,
$$
$$
	k(s) = \frac{\sqrt{s-4 m_{\pi}^2}}{2}\,, \quad f(s, m_V, \Gamma_V) = \Gamma_V \frac{m_V^2}{k(
    m_V^2)^3} [k^2(s) (h(s) - h(m_V^2)) - (s - m_V^2) k^2(m_V^2) h'(m_V^2)],
$$
$$
	h(s) = \frac{2}{\pi} \frac{k(s)}{\sqrt{s}} \ln \left( \frac{\sqrt{s} + 2 m_{\pi}}{2 m_{\pi}} \right), \quad 
	h'(mV^2) = h'(s)\left|_{s=mV^2} \right.,\quad
	F^\text{BW}(0, m_V, \Gamma_V)=1 \, \text{automatically}.
$$
The full pion form-factor with resonances~$\rho$, $\omega$ and~$\rho'$ has the following expression:
$$
 \label{eq:4res}
  F_{\pi}\left(s, \{\rho,\omega, \rho' \}\}\right)
  =\frac{F_\pi(s,\{\rho,\omega\})
      + \alpha_{\rho'} F^\text{BW}\left(s,m_{\rho'},\Gamma_{\rho'}\right)}
       {1+\alpha_{\rho'}}\,,~~~
$$
$$
 \text{where} \quad F_{\pi}\left(s, \{\rho,\omega\}\right)
  = F^\text{BW}(s,m_{\rho},\Gamma_{\rho})\,
     \frac{1+\alpha_{\omega} F^\text{BW}(s,m_{\omega},\Gamma_{\omega})}
          {1+\alpha_{\omega}}\,.
$$

Performing the $\chi^2$-minimisation we get the fit of the experimental data on~$|F_{\pi}|^2(s)$ taken from work~\cite{Barkov85} characterised by
$\chi^2_\text{b.f.}=0.82$
and the following resonances parameters values\footnote{We use values of~$m_{\rho}$, $m_{\omega}$ and~$\Gamma_{\omega}$ taken from PDG~\cite{PDG}} :
\begin{eqnarray}
   m_{\rho}&=&0.770~\text{GeV~(PDG)}\,,\quad
   \Gamma_{\rho}\,=\,0.149 \pm 0.008~\text{GeV}\,;\nonumber\\
   m_{\omega}&=&0.782~\text{GeV~(PDG)}\,,\quad
   \Gamma_{\omega}\,=\,0.009~\text{GeV~(PDG)}\,,\quad
   \alpha_{\omega}\,=\,0.0021 \pm 0.0017\,;\nonumber\\
   m_{\rho'}&=&1.354 \pm 0.116~\text{GeV}\,,\quad
   \Gamma_{\rho'}\,=\,0.344 \pm 0.157~\text{GeV}\,,\quad
   \alpha_{\rho'}\,=\,-0.089 \pm 0.024\,;\nonumber\\
	 d\,&=&\,0.384 \pm 0.200~\text{GeV}\,.\nonumber
\end{eqnarray}

For cross-sections of the processes  
$e^+e^-\to2\pi^+2\pi^-$,
$e^+e^-\to\pi^+\pi^-2\pi^0$,
$e^+e^-\to2\pi^+2\pi^-2\pi^0$,
and
$e^+e^-\to3\pi^+3\pi^-$
the description in the form of sum of three Gaussian curves, describing wide resonances, is assumed: 
\begin{eqnarray*}
 \label{eq:Gauss.3R}
   \sigma\left(s, \{M_{i},\sigma_{i},\alpha_{i}\}\right)
    = \sum_{i=1}^3\frac{\alpha_i^2}{\alpha_1^2+\alpha_2^2+\alpha_3^2}\,
       e^{-(\sqrt{s}-M_i)^2/(2\sigma_i^2)}\,.
\end{eqnarray*}
The results are shown in table~\ref{tab-1}.
\begin{table}
\centering
\caption{The fitting results for particular $e^+e^-$-annihilation channels}
\label{tab-1}       
\begin{tabular}{r|llll}
\hline
 $e^+e^-\to2\pi^+2\pi^- \quad i$ &$M_i$, GeV & $\sigma_i$, GeV & $\alpha_i$ \\\hline
1&1.289 $\pm$ 0.034 & 0.213 $\pm$ 0.028 & 0.462 $\pm$ 0.058 \\
$\chi^2_\text{b.f.}=0.97\quad$
2&1.573 $\pm$ 0.017 & 0.227 $\pm$ 0.011 & 1.522 $\pm$ 0.121 \\
3&2.524 $\pm$ 0.070 & 0.356 $\pm$ 0.037 & 2.427 $\pm$ 0.791 \\\hline
\hline
 $e^+e^-\to\pi^+\pi^-2\pi^0 \quad i$ &$M_i$, GeV & $\sigma_i$, GeV & $\alpha_i$  \\\hline
1&$1.304 \pm 0.250$ & $0.258 \pm 0.066$ & $1.024 \pm 1.064$ \\
$\chi^2_\text{b.f.}=1.12\quad$
2&$1.748 \pm 0.411$ & $0.283 \pm 0.052$ & $1.333 \pm 0.275$ \\
3&$2.322 \pm 0.201$ & $0.194 \pm 1.027$ & $0.817 \pm 2.826$ \\\hline
\hline
 $e^+e^-\to 3\pi^+3\pi^- \quad i$ &$M_i$, GeV & $\sigma_i$, GeV & $\alpha_i$  \\\hline
1&$1.811\pm 0.027$ & $0.091\pm  0.023$ & $0.183\pm 0.040$ \\
$\chi^2_\text{b.f.}=0.58\quad$
2&$1.897\pm 0.034$ & $0.237\pm 0.027$ & $\,0.291\pm 0.029$ \\
3&$2.247\pm 0.042$ & $0.134\pm 0.264$ & $0.956\pm 0.534$ \\\hline
\hline
 $e^+e^-\to 2\pi^+2\pi^-2\pi^0 \quad i$ &$M_i$, GeV & $\sigma_i$, GeV & $\alpha_i$  \\\hline
1&$1.740\pm 0.023$ & $0.115\pm 0.019$ & $0.529\pm 0.082$ \\
$\chi^2_\text{b.f.}=0.75\quad$
2&$2.027\pm 0.080$ & $0.287\pm 0.062$ & $0.362\pm 0.082$ \\
3&$2.274\pm 0.042$ & $0.248\pm 0.026$ & $1.051\pm 0.255$ \\\hline
\end{tabular}
\end{table}

\section{$R$-ratio, Adler function and sum rule  
\label{sec-RD}
}


The full $R$-ratio is the sum of $R$-ratios of particular processes.
$ 
 R(s)
   = \sum\limits_{i} \cfrac{\sigma_{e^+e^-\to\text{hadrons of type $i$}}(s)}
             {\sigma_{e^+e^-\to \mu^+\mu^-}(s)}\,.
$
\begin{figure*}
	\centering
				\includegraphics[width=12cm,clip]{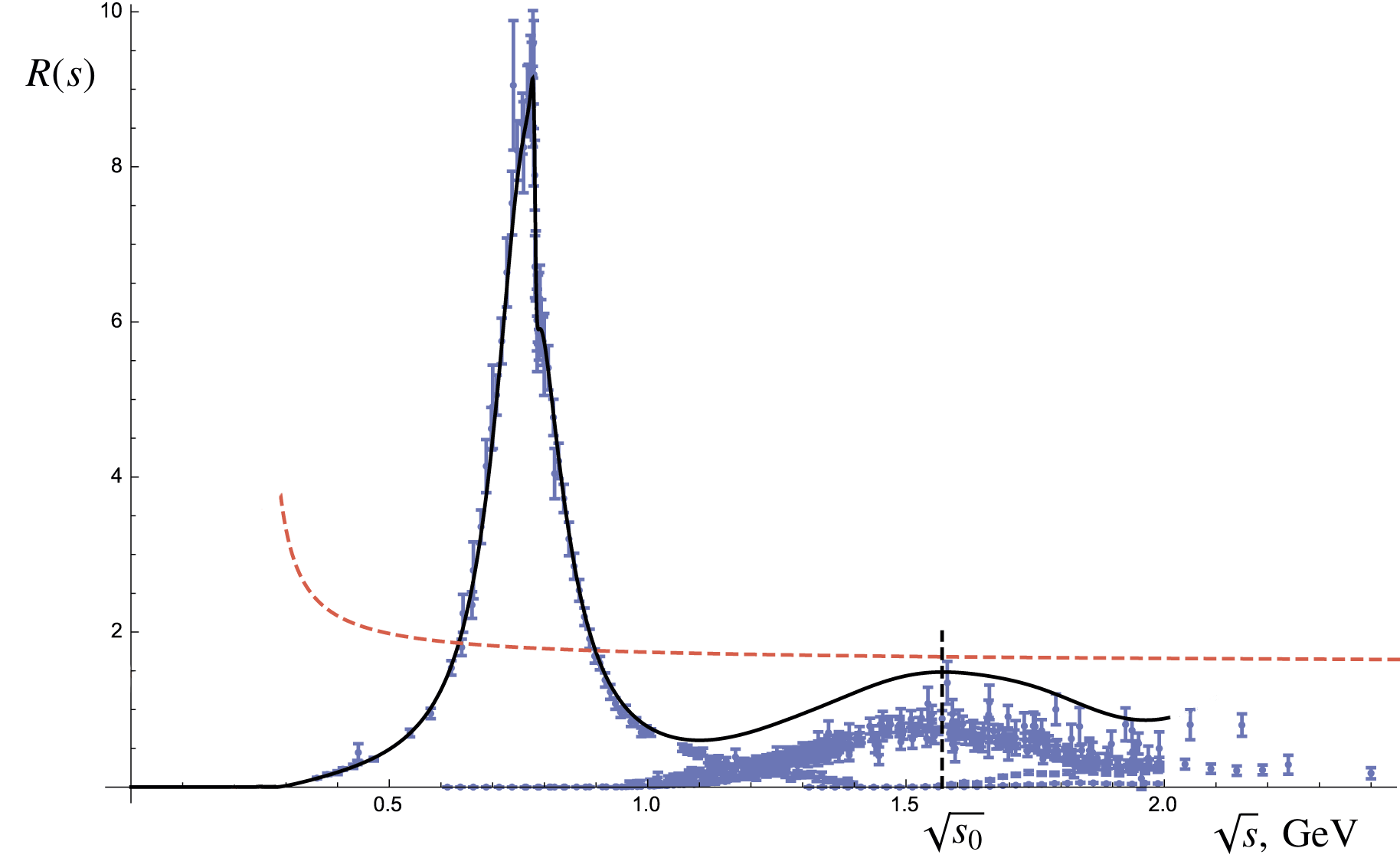}
		\vspace*{0cm}       
		\caption{The full~$R$-ratio in dependence on energy~$\sqrt{s}$ at~$\sqrt{s}\leq2$~GeV and the theoretical representation of $R$-ratio. The data on particular channels of $e^+e^-$-annihilation are presented by blue dots with errors.}
	\label{fig:Rs0}      
\end{figure*}

Extracting the current with isospin~$I=1$ ($j_{\mu}=\frac 1 2 (\bar u \gamma_{\mu} u-\bar d \gamma_{\mu} d)$) from full electromagnetic current, we obtain the formula for $D$-function with isospin $I=1$ in the framework of OPE (for example, see work~\cite{SVZ}, eq.~3.4):
\begin{eqnarray}
\label{eq:Adler.PT-OPE}
  D_\text{PT+OPE}(Q^2)
   = \frac 3 2 \,
      \left[1 + \frac{\alpha_s(Q^2)}{\pi}
            + \sum_{n\geq 1} \Gamma(n)\,\frac{a_{2n}}{Q^{2n}}
      \right],
\end{eqnarray}
where
~$a_{2n}$ are OPE coefficients, we take into account three first power corrections.

The another form of~$D$-function, dispersional, is the following:
\begin{eqnarray}
 \label{eq:D.exp}
  D_\text{exp}(Q^2)=
	Q^2 \int_{4\,m_\pi^2}^{\infty}\frac{R_\text{exp-th}(s)\,ds}{(s+Q^2)^2}
   = Q^2 \int_{4\,m_\pi^2}^{s_0}\frac{R_\text{exp}(s)\,ds}{(s+Q^2)^2}
   + Q^2 \int_{s_0}^\infty\frac{R_\text{th}(s)\,ds}{(s+Q^2)^2}\,,
\end{eqnarray}
$$
\text{where} \quad R_\text{exp-th}(s) = R_\text{exp}(s)\,\theta(s<s_0) + R_\text{th}(s)\,\theta(s>s_0),\quad
  R_\text{th}(s)= \frac{3}{2}\,\left[1+\frac{\alpha_s(s)}{\pi}\right];
$$
$R_\text{exp}(s)$  is our fitting result (shown by black solid curve on Fig.~\ref{fig:Rs0}) and~$R_\text{th}(s)$ is the one-loop approximation of perturbative QCD (shown by red dashed curve on Fig.~\ref{fig:Rs0}).

The continuum threshold $\sqrt{s_0} = 1.57$~GeV is chosen to guarantee that $R_\text{exp}(s)$ and $R_\text{th}(s)$ and even their first derivatives take similar values (under the statistical uncertainties of experimental data), as it is seen on Fig.~\ref{fig:Rs0} (vertical dashed line).
The function~$R_\text{exp}(s)$ decreases above~$\sqrt{s_0} = 1.57$~GeV, which is explained by the absence of the data on~$e^+e^-$-annihilation to~$8\pi$ channels.




Equating the both forms of Adler function: by OPE~(\ref{eq:Adler.PT-OPE}) and dispersional~(\ref{eq:D.exp}), and applying BT, we get the sum rule:
\begin{eqnarray}
\label{eq:SumRule}
		\Phi_\text{exp}(M^2) = \Phi_{\text{PT+OPE}}(M^2),
\end{eqnarray}
\begin{eqnarray*}
 \label{eq:Ad.Bor.Exp}
   \text{where}\quad \Phi_\text{exp}(M^2)
    = \int_0^{\infty}\!\!
         R_\text{exp-th}(s)\,\left(1-\frac{s}{M^2}\right)\,e^{-s/M^2}\,
         \frac{ds}{M^2},
\end{eqnarray*}
 \begin{equation*}
 \label{eq:Ad.Bor.PT.OPE}
   \Phi_{\text{PT+OPE}}(M^2) = 
			\frac 3 2 \frac{\hat{B}_{Q^2\to M^2}[ \alpha_s(Q^2)]}{\pi}
				  + \frac 3 2 \left( \frac{C_2}{M^{2}}
          + \frac{C_4}{M^{4}}
          + \frac{C_6}{M^{6}} \right), \quad C_{2n}=a_{2n},
  \end{equation*}
\begin{eqnarray*}
 \label{eq:Bor.abar}
		\text{and}\quad \hat{B}_{Q^2\to M^2}\left[\alpha_s(Q^2)\right] 
   = \frac{4\pi}{b_0}\, \left[\frac{1}{M^2}\,\int \limits_0^{\infty}\!\! \frac{e^{-s/M^2}d s}{\ln^2\left(s/\Lambda^2\right)+\pi^2}
   + \frac{\Lambda^2}{M^2}\,e^{\Lambda^2/M^2}\right]\,,\quad b_0=11-2N_f/3,
\end{eqnarray*}
$\Lambda=\Lambda_{\text{QCD}}=0.25$~GeV is the QCD scale parameter.
$M^2$ is the Borel mass, appearing in equation after Borel transform. 

\section{The sum rule analysis}
\label{sec-Analysis}
Finally, let us turn to the discussion of obtained sum rule~(\ref{eq:SumRule}).
The Borel mass~$M^2$ is varied in interval~$0.75 \div 4$~GeV$^2$ which is divided to 20 points, then the coefficients~$C_2$ and~$C_4$ are determined by~$\chi^2$-minimisation, and $C_6 =-\frac{448\,\pi^3}{27}\,\alpha_s \left\langle \bar{q}q\right\rangle^2 =-0.121$~GeV$^6$, which can be expressed in terms of quark condensate~\cite{SVZ}, is fixed ~\cite{GeshIoffeZyabl}.
The coefficient~$C_4$ can be expressed in terms of gluon condensate
	$	C_4 = \frac{2\pi^2}{3}\,\left\langle \frac{\alpha_s \, GG}{\pi}\right\rangle$~\cite{SVZtf79}. 

\begin{figure}[h]\center
	\includegraphics[height=7cm]{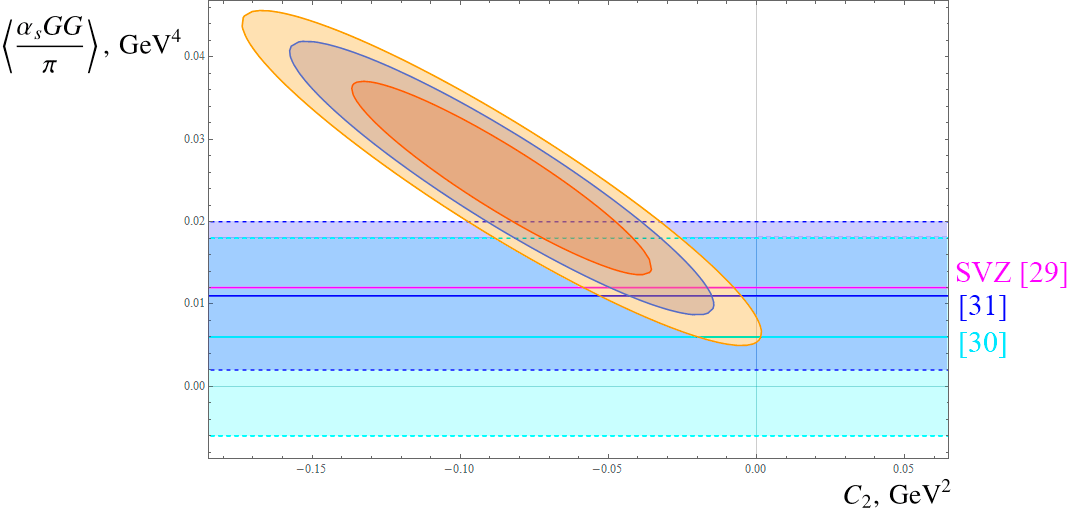}
	\caption{
	The allowed regions for~$C_2$ and gluon condensate at 1, 2 and 3~$\sigma$-level, determined by $\chi^2 \leq \chi_{\text{min}}^2+1$ (red), $\chi^2 \leq \chi_{\text{min}}^2+2$ (blue) and $\chi^2 \leq \chi_{\text{min}}^2+3$ (yellow). 
	The regions of existing data on gluon condensate are shown for comparison.
	Central values are marked by solid horizontal lines,  errors ranges are marked by dashed lines.}
\label{fig:Chi2Region}
\end{figure}

Our result is shown on Fig.~\ref{fig:Chi2Region}; the regions corresponding to 1, 2 and 3~$\sigma$-levels are marked in red, blue and yellow.
The minimal value is~$\chi_{\text{min}}^2=0.483$, 
and corresponding gluon condensate and~$C_2$ are found to be:~$\left\langle (\alpha_s/\pi) \,GG\right\rangle = 0.025\,\text{GeV}^4$, $C_2 = -0.086\,\text{GeV}^2$. The allowed intervals for~$C_2$ and gluon condensate at one~$\sigma$-level are: 
$$
C_2=-0.086 \pm 0.050\,\text{GeV}^2, \quad
\left\langle \frac{\alpha_s  GG}{\pi}\right\rangle=0.025 \pm 0.012\,\text{GeV}^4.
$$

At one $\sigma$-level our results do not contradict with the well-known results~\cite{SVZ, GeshIoffeZyabl, IoffeGG}, but at the same time~$C_2$ is not compatible with zero. 
One can see that~$C_2=0$ is possible at 3~$\sigma$-level.

Furthermore, by the form of regions corresponding our results~(see Fig.~\ref{fig:Chi2Region}) we can suppose that there is some (anti)correlation between~$C_2$ and~$C_4$. 
A rough estimation of this connection is expressed by formula: 
$$
	C_2 \approx -5\,\text{GeV}^{-2}\, \left\langle \frac{\alpha_s  GG}{\pi}\right\rangle +0.025\,\text{GeV}^2.
$$
One can compare the received values with the existing values of the gluon condensate
~\cite{SVZ,IoffeGG,GeshIoffeZyabl}.


\section{Conclusions and outlook}
\label{sec-Results}
The resonance contribution fitting model is developed,
the Adler function with Borelization is obtained and its precise numerical analysis is performed.

$C_2$ has negative sign and is compatible with zero only at 3 $\sigma$-level. At 1 $\sigma$-level:
$C_2=-0.086 \pm 0.050$~GeV$^2$, 
$\left\langle \alpha_s/\pi \, GG\right\rangle=0.025 \pm 0.012$~GeV$^4$. 

Strong (anti)correlation between short strings and local gluon condensate is found:
$$
C_2 \approx -5\,\text{GeV}^{-2}\, \left\langle \frac{\alpha_s  GG}{\pi}\right\rangle +0.025\,\text{GeV}^2.
$$






We plan to perform an analogous analysis in the framework of the analytic perturbation theory~\cite{Shirkov:1997wi, Bakulev:2005gw, Khandramai:2011zd}.

\textit{Acknowlegements. 
\\
We dedicate this paper to the memory of Alexander Petrovich Bakulev (1956-2012) with whom we started this work. 
We are grateful to A.~L.~Kataev, S.~V.~Mikhailov, 
O.~P.~Solovtsova for useful discussions and comments. M.~K. is thankful to Yu.~L.~Ryzhykau for valuable advices and assistance in numerical analysis. The work is supported in part by RFBR Grant-140100647.
}

\begin{thebibliography}{}
%
%
\bibitem{SVZtf79}
M. A. Shifman, A. I. Vainshtein, V.I. Zakharov. 
Nucl. Phys. \textbf{B147}, 385-447 (1979).

\bibitem{ShortString}
K.G. Chetyrkin, S. Narison, V.I. Zakharov. 
Nucl.Phys. \textbf{B550}, 353-374 (1999).

\bibitem{ZakharovOther}
S. Narison, V.I. Zakharov. 
Phys. Lett.
\textbf{B679}, 355-361(2009).

\bibitem{BodDominEidel2012}
S. Bodenstein, C.A. Dominguez, S.I. Eidelman, H. Spiesberger, K. Schilcher. 
JHEP 1201 \textbf{039} (2012). 

\bibitem{Eichten:1974af} 
  E.~Eichten, K.~Gottfried, T.~Kinoshita, J.~B.~Kogut, K.~D.~Lane and T.~M.~Yan,
  Phys.\ Rev.\ Lett.\  {\bf 34}, 369 (1975)
  Erratum: [Phys.\ Rev.\ Lett.\  {\bf 36}, 1276 (1976)].

\bibitem{Kataev:2000wn} 
  A.~L.~Kataev, G.~Parente and A.~V.~Sidorov,
  Fiz.\ Elem.\ Chast.\ Atom.\ Yadra {\bf 31P7B}, 29 (2000)
  [Phys.\ Part.\ Nucl.\  {\bf 31}, no. 7B, 29 (2000)].

\bibitem{SPIESBERGER:2013jpa} 
  H.~Spiesberger,
  Mod.\ Phys.\ Lett.\ A {\bf 28}, 1360009 (2013).
	
\bibitem{Barkov85}
L.~M. Barkov 
et~al.
Nucl. Phys.
\textbf{B256}, 365--384 (1985).

\bibitem{M3N79}
G.~Cosme, B.~Dudelzak, B.~Grelaud, B.~Jean-Marie, S.~Jullian et~al. 
Nucl. Phys. \textbf{B152}, 215-231 (1979).

\bibitem{GG280}
C.~Bacci, R.~Baldini-Celio, G.~Battistoni, G.~Capon, R.~Del~Fabbro et~al.
Phys. Lett. \textbf{B95}, 139-142  (1980).

\bibitem{DM182}
A.~Cordier, D.~Bisello, J.~Bizot, J.~Buon, B.~Delcourt et~al.
Phys. Lett. \textbf{B109}, 129-132 (1982).

\bibitem{CMD88}
L.~M. Barkov, I.~B. Vasserman, P.~V. Vorobev, P.~M. Ivanov, G.~Y. Kezerashvili et~al.
Sov. J. Nucl. Phys.
\textbf{47}, 248--252 (1988).

\bibitem{OLYA88}
L.~M. Kurdadze
et~al.
JETP Lett.
\textbf{47}, 512--515 (1988).

\bibitem{DM291}
D.~Bisello et~al.
Nucl. Phys. Proc. Suppl.
\textbf{21}, 111--117 (1991).

\bibitem{SND91}
S.~I. Dolinsky, V.~P. Druzhinin, M.~S. Dubrovin, V.~B. Golubev, V.~N. Ivanchenko et~al.
Phys.Rept.
\textbf{202}, 99--170 (1991).

\bibitem{CMD299}
R.~R. Akhmetshin et~al.
Phys. Lett.
\textbf{B475}, 190--197 (2000).

\bibitem{SND03}
M.~N. Achasov, K.~I. Beloborodov, A.~V. Berdyugin, A.~G. Bogdanchikov, A.~V. Bozhenok et~al.
J. Exp. Theor. Phys.
\textbf{96}, 789--800 (2003).

\bibitem{CMD204}
R.~R. Akhmetshin et~al.
Phys. Lett.
\textbf{B595}, 101--108 (2004).

\bibitem{BaBar05}
B.~Aubert et~al.
Phys. Rev.
\textbf{D71}, 052001 (2005).

\bibitem{All-4pi0}
M.~R.~Whalley.
J. Phys. G: Nucl. Part. Phys. 
\textbf{29},A1-A133 (2003).

\bibitem{OLYA4pi0}
L.~M.~Kurdadze et al.
J. Exp. Theor. Phys. Lett. 
\textbf{43}, 643-645 (1986).

\bibitem{CMD24pi0}
R.~R.~Akhmetshin et al.
Phys. Lett.
\textbf{B466}, 392-402 (1999).

\bibitem{SND4pi0}
M.~N.~Achasov et al.
Preprint BUDKER-INP-2001-34 (2001).

\bibitem{ORSAY-DCI-DM2-4pi0}
B.~Bisello et al.
\textit{DM2 results on e+ e- annihilation into multi - hadrons in the 1350~MeV - 2400~MeV energy range.
High-energy physics. Proceedings.} (25th International Conference, Singapore, August 2-8, 1990) Vol. I+II.
Preprint LAL-90-35 (1990).

\bibitem{FRASCATI-ADONE-4pi0}
C.~Bacci et al.
Nucl. Phys. \textbf{B184}, 31-39 (1980).

\bibitem{BaBar06}
B.~Aubert et~al.
Phys. Rev.
\textbf{D73}, 052003 (2006).

\bibitem{GouSak68}
G.~J. Gounaris, J.~J. Sakurai
Phys. Rev. Lett.
\textbf{21},244--247 (1968).

\bibitem{PDG}
K.~A.~Olive et al.
Particle Data Group.
Chin. Phys.
\textbf{C38},1025--1034 (2014 and 2015 update).




\bibitem{SVZ}
M.~A. Shifman, A.~I. Vainshtein, V.~I. Zakharov
Nucl. Phys. \textbf{B147}, 448-518  (1979).

\bibitem{GeshIoffeZyabl}
B. V. Geshkenbein, B. L. Ioffe, and K. N. Zyablyuk. 
Phys. Rev. \textbf{D64}, 093009 (2001).

\bibitem{IoffeGG}
B. L. Ioffe, K. N. Zyablyuk. 
Eur. Phys. J. \textbf{C27},229-241 (2003).

\bibitem{Shirkov:1997wi} 
  D.~V.~Shirkov and I.~L.~Solovtsov,
  Phys.\ Rev.\ Lett.\  {\bf 79}, 1209 (1997),
  Theor.\ Math.\ Phys.\  {\bf 150}, 132 (2007).

\bibitem{Bakulev:2005gw} 
  A.~P.~Bakulev, S.~V.~Mikhailov and N.~G.~Stefanis,
  Phys.\ Rev.\ D {\bf 72}, 074014 (2005)
  [Phys.\ Rev.\ D {\bf 72}, 119908 (2005)].

\bibitem{Khandramai:2011zd} 
  V.~L.~Khandramai, R.~S.~Pasechnik, D.~V.~Shirkov, O.~P.~Solovtsova and O.~V.~Teryaev,
  Phys.\ Lett.\ B {\bf 706}, 340 (2012).


\end{thebibliography}
%
%

\end{document}